\documentclass[12pt,preprintnumbers,nofootinbib,letterpaper]{article}
\pdfoutput=1
\usepackage{amsmath,amssymb,amsfonts,cite,xcolor,epsf,epsfig,epstopdf,graphics,booktabs,graphicx,mathtools,subcaption,physics,verbatim}
\def\thefootnote{*\arabic{footnote}}
\definecolor{ultramarine}{rgb}{0.07, 0.04, 0.56}
\definecolor{cadmiumgreen}{rgb}{0.0, 0.42, 0.24}
\definecolor{indigo(dye)}{rgb}{0.0, 0.25, 0.42}
\usepackage[linktocpage=true,breaklinks]{hyperref}
\hypersetup{
	colorlinks=true,
	citecolor=ultramarine,
	linkcolor=cadmiumgreen,
	urlcolor=indigo(dye),
}

\usepackage[utf8]{inputenc}

\numberwithin{equation}{section}

\usepackage{array}
\newcolumntype{P}[1]{>{\centering\arraybackslash}p{#1}}

\usepackage[shortlabels]{enumitem}

\usepackage{dcolumn}
\newcolumntype{M}[1]{>{\centering\arraybackslash}m{#1}}
\newcolumntype{N}{@{}m{0pt}@{}}

\usepackage{amsmath}
\usepackage{ulem}

\usepackage[utf8]{inputenc}
\usepackage[T1]{fontenc}
\usepackage{lmodern}
\usepackage{geometry}
\usepackage{amsmath,amssymb,amsthm,mathtools}
\usepackage{bm}
\usepackage{microtype}
\usepackage{physics}

\newcommand{\Mpl}{M_{\rm Pl}}

\newcommand{\D}{{\rm d}}

\newcommand{\be}{\begin{equation}}  
	\newcommand{\ee}{\end{equation}}

\usepackage{tikz}

\begin{document}
	

	\begin{center}
		
		\def\thefootnote{\fnsymbol{footnote}}
		
		\vspace*{1.5cm}
		{\Large {\bf Kinematical correlations via $\kappa$-Poincar\'e coproducts}}
		\\[1cm]
		
		{Mohammad Ali Gorji$^{1}$ and Babak Vakili$^{2}$}
\\[.7cm]

{\small\textit{$^{1}$Cosmology, Gravity, and Astroparticle Physics Group, Center for Theoretical Physics of the Universe, Institute for Basic Science (IBS), Daejeon, 34126, Korea}}\\
\vspace{0.25cm}
{\small\textit{$^{2}$Department of Physics, CT.C., Islamic Azad University, Tehran, Iran}}

\end{center}

\vspace{1cm}

\hrule \vspace{0.5cm}

\begin{abstract}
We study a kinematical consequence of the Hopf-algebraic momentum composition law in $\kappa$-Minkowski spacetime. The same curved momentum space can be described in different coordinates. In the bicrossproduct basis the ordered-plane-wave labels are the translation-generator eigenvalues, so the relevant map is one-to-one. In the classical basis, instead, the translation eigenvalues $P_\mu$ are nonlinearly related to the ordered-plane-wave labels $p_\mu$. This relation can fail to be globally one-to-one in a high-momentum region. When a given classical-basis four-momentum admits more than one real auxiliary preimage, the branch-sensitive quantity $P_+\equiv P_0+P_4=\kappa e^{p_0/\kappa}$ enters the coproduct and resolves the branches in two-particle states. Imposing the vanishing total-momentum constraint therefore gives branch-dependent $\kappa$-deformed back-to-back momentum correlations. In a single-branch regime this is just a deformed correlated product, while in a multibranch regime a state specified only by $P_\mu$ can be expanded into distinct auxiliary branches. If $P_\mu$ are taken as the directly meaningful momenta, the physical content is the resulting deformed correlation pattern. If the auxiliary variables $p_\mu$ are assigned operational meaning, the same constrained state can be interpreted as a superposition over different auxiliary branches. We also compare this structure with standard regular self-adjoint nonrelativistic minimal-length models and find no analogous smooth local two-real-branch inversion on their physical domains.
\end{abstract}

\vspace{0.5cm}

\hrule
\def\thefootnote{\arabic{footnote}}
\setcounter{footnote}{0}

\newpage

\section{Introduction}\label{sec:introduction}
	
	A recurring lesson from many approaches to quantum gravity is that a new ultraviolet
(UV) scale should enter the description of spacetime and matter.
At a purely dimensional level, combining $\hbar$, $c$, and $G$ singles out the Planck energy
(or equivalently the Planck length/time), suggesting that quantum-gravitational effects may be organized
as deformations controlled by an invariant high-energy scale.
This expectation becomes particularly relevant in regimes where one can neglect strong curvature
effects and also work with semiclassical matter, yet the UV scale still persists as a structural
ingredient of the effective theory.
Motivated by this idea, a variety of frameworks, including noncommutative geometry,
polymer quantization, generalized uncertainty principles, and related effective models, have been developed
to parameterize potential Planckian departures from standard kinematics
\cite{Kempf:1994su, Kempf:1996nk, Amelino-Camelia:2000cpa, Hossenfelder:2012jw, Nozari:2012gd, Tawfik:2015rva, Snyder:1946qz, Doplicher:1994tu}.

A particularly promising and conceptually sharp direction is doubly special relativity (DSR),
in which special relativity is extended so that, in addition to the speed of light,
there is a second observer-independent scale, typically taken to be a UV (Planckian) energy scale.
This allows one to preserve the relativity principle while accommodating deformed
kinematics at high energies \cite{Amelino-Camelia:2000stu, Kowalski-Glikman:2004fsz, Freidel:2003sp}.
One of the simplest and most studied realizations of this idea is $\kappa$-Minkowski spacetime,
characterized by Lie-type noncommutative coordinates and an observer-independent deformation scale $\kappa$,
often associated with the Planck energy \cite{Amelino-Camelia:2000stu, Kowalski-Glikman:2004fsz, Freidel:2003sp}.
Depending on the physical implementation and on how operational notions of measurement are modeled,
the deformation scale $\kappa$ can be interpreted in different but related ways, for example as an
invariant UV energy scale (with an associated length $\ell_\kappa\sim 1/\kappa$), or effectively as a
maximal energy/momentum scale or a minimal resolvable length.

The symmetry structure underlying $\kappa$-Minkowski spacetime is provided by the $\kappa$-Poincar\'e algebra,
a deformation of the ordinary Poincar\'e algebra with a nontrivial Hopf-algebraic coproduct
\cite{Kowalski-Glikman:2001gmh, Lukierski:1992dt, Majid:1994cy, Agostini:2003vg}.
In this framework, the momentum sector can be parametrized in different but equivalent ways, and the deformed coproduct encodes a nonlinear law for momentum composition on composite systems.
This feature reflects the curved geometry of momentum space and plays a central role in ensuring
the consistency of relativistic transformations when the UV scale $\kappa$ is required to be observer independent.
In particular, the nonlinear composition of momenta underlies the relativity of locality viewpoint and related
geometrical formulations of DSR kinematics \cite{Amelino-Camelia:2011lvm, Amelino-Camelia:2011hjg}.

Quantum field theories on $\kappa$-Minkowski spacetime are conveniently formulated in terms of ordered plane waves,
which provide a faithful representation of fields and lead naturally to a nonlinear relation between ordered-plane-wave labels and classical-basis translation-generator eigenvalues.
As a consequence, a single classical-basis four-momentum can admit multiple preimages in auxiliary momentum space.
While this multivalued structure has been studied in several discussions of deformed dispersion relations, momentum-space geometry, and relative locality \cite{Amelino-Camelia:2002cqb, Lukierski:1993wx, Amelino-Camelia:2011lvm, Amelino-Camelia:2011hjg, Arzano:2022ewc},
its implications for purely kinematical two-particle correlations have remained largely unexplored.

The rest of the paper is organized as follows. In Sec.~\ref{sec:kappa_poincare} we summarize the
$\kappa$-Minkowski plane-wave framework, review both the bicrossproduct and classical-basis presentations of the $\kappa$-Poincar\'e translation sector (including their nonprimitive coproduct structures), and then focus on the nonlinear map relating the classical-basis variables $P_\mu$ to the ordered-plane-wave variables $p_\mu$.
In Sec.~\ref{sec:coproduct_entanglement} we impose the classical-basis coproduct total-momentum constraint and show that,
in a high-momentum regime where the inverse map admits two real branches, the constrained two-particle
state exhibits $\kappa$-deformed momentum correlations and admits a nontrivial auxiliary branch expansion.
We also discuss the interpretation of the two viewpoints on $(P_\mu,p_\mu)$ and briefly comment on what
would be required to access the effect phenomenologically. In Sec.~\ref{sec:GUP_relation} we contrast this with standard nonrelativistic
minimal-length scenarios, such as polymer quantum mechanics and generalized uncertainty principle models, and show that they do not support an analogous smooth local two-branch inversion on their physically
admissible domains. Sec.~\ref{sec:summary} is devoted to the summary and conclusion. Finally, in
Appendix~\ref{sec:lorentz_covariance} we prove that the construction is covariant under $\kappa$-deformed
Lorentz transformations in the classical basis.

\section{$\kappa$-Poincar\'e kinematics}
\label{sec:kappa_poincare}

A common way to incorporate the existence of a fundamental energy or length scale into relativistic physics
is to relax the assumption of commuting spacetime coordinates.
One of the most studied examples of such noncommutative geometries
is the $\kappa$-Minkowski spacetime, defined by the Lie-algebra-type
commutation relations \cite{Majid:1994cy}
\begin{equation}\label{eq:kappa_minkowski_coords}
	[x^0, x^i] = \frac{i}{\kappa} x^i, 
	\qquad
	[x^i, x^j] = 0 .
\end{equation}
Here $\kappa$ is a deformation scale typically associated with the Planck energy.
Moreover, within a given representation of the $\kappa$-Minkowski coordinate algebra,
the noncommutativity in \eqref{eq:kappa_minkowski_coords} implies that temporal and spatial coordinates
cannot be simultaneously specified with arbitrary sharpness. Depending on the chosen realization of the
associated phase space and on how measurements are implemented, this $\kappa$-controlled fuzziness is
often described in terms of an effective minimal time/length scale or, dually, a maximal energy/momentum scale.

Fields on $\kappa$-Minkowski spacetime can be described by a Fourier-type expansion in terms of noncommutative plane waves. In ordinary Minkowski space one uses $e^{ip_\mu x^\mu}=e^{i(p_i x^i+p_0 x^0)}$, but in the noncommutative coordinate algebra \eqref{eq:kappa_minkowski_coords} the time coordinate $x^0$ does not commute with the spatial coordinates $x^i$, so the exponential $e^{ip_\mu x^\mu}$ must be defined with a definite ordering. This naturally leads to ordered plane waves and, via their multiplication, to a nonlinear composition rule for the momentum labels. A common choice (time-to-the-right ordering) is \cite{Lukierski:1992dt}
\begin{equation}\label{eq:ordered_plane_waves}
	e_p(x)\equiv e^{i p_i x^i}\,e^{i p_0 x^0},
\end{equation}
where the ordering prescription reflects the underlying noncommutative structure. These ordered plane waves provide a faithful representation
of functions on $\kappa$-Minkowski spacetime and play the role
of generalized plane waves adapted to its noncommutative structure
\cite{Majid:1994cy, Agostini:2003vg}. The product of two plane waves induces a deformed addition law for the ordered-plane-wave labels
\begin{equation}\label{eq:ep-eq}
	e_p(x)\, e_q(x)
	= e_{p \oplus q}(x),
\end{equation}
with\footnote{The deformed momentum addition law follows directly from the noncommutative structure of $\kappa$-Minkowski spacetime together with the ordering choice \eqref{eq:ordered_plane_waves}. Using the first commutation relation in \eqref{eq:kappa_minkowski_coords},
	one finds
	$e^{i p_0 x^0} x^i e^{-i p_0 x^0}=e^{-p_0/\kappa}x^i$.
	As a result, the product of ordered plane waves
	$e_p(x)=e^{i p_i x^i}e^{i p_0 x^0}$
	satisfies
	$e_p(x)e_q(x)
	= e^{i (p_i+e^{-p_0/\kappa}q_i)x^i}e^{i(p_0+q_0)x^0}$,
	which induces the nonlinear composition rule
	\eqref{eq:deformed_addition}.}
\begin{equation}\label{eq:deformed_addition}
	(p \oplus q)_0 = p_0 + q_0,
	\qquad
	(p \oplus q)_i = p_i + e^{-p_0/\kappa} q_i .
\end{equation}
In the commutative ($\kappa\to\infty$) limit one recovers the usual Minkowski plane waves
$e^{ip_\mu x^\mu}$ and the standard linear addition of momenta $(p\oplus q)_\mu\rightarrow p_\mu+q_\mu$. The nonlinearity in \eqref{eq:deformed_addition} therefore measures the departure from the ordinary vector-space structure of momentum addition induced by the $\kappa$-Minkowski ordering.

At this stage it is useful to separate two equivalent but differently parametrized presentations of the same curved momentum space. The first uses directly the ordered-plane-wave labels $p_\mu$, which are naturally associated with the bicrossproduct basis. The second uses the classical-basis variables $P_\mu$, nonlinearly related to $p_\mu$. Since the rest of the paper will be formulated in the latter presentation, we first summarize the bicrossproduct basis and then pass to the classical basis.

\subsection{Bicrossproduct basis}
\label{subsec:bicrossproduct_basis}

In the bicrossproduct basis one identifies the translation-generator eigenvalues with the ordered-plane-wave labels themselves \cite{Majid:1994cy}. Thus the action of the translation generators on ordered plane waves is
\begin{equation}\label{eq:bicross_action_on_plane_waves}
	p_\mu \triangleright e_p(x)=p_\mu\,e_p(x).
\end{equation}
The ordered-plane-wave composition law \eqref{eq:deformed_addition} is then encoded directly in the coproduct
\begin{align}
	\Delta(p_0) &= p_0 \otimes 1 + 1 \otimes p_0,
	\label{eq:bicross_coproduct_energy}
	\\
	\Delta(p_i) &= p_i \otimes 1 + e^{-p_0/\kappa} \otimes p_i ,
	\label{eq:bicross_coproduct_momentum}
\end{align}
so that on product eigenstates one recovers
\begin{align}
	(p \oplus q)_0 &= p_0+q_0,
	\label{eq:bicross_composition_energy}
	\\
	(p \oplus q)_i &= p_i+e^{-p_0/\kappa}q_i.
	\label{eq:bicross_composition_momentum}
\end{align}
The corresponding antipode, defined by the condition $p \oplus S(p)=0$, is
\begin{align}
	S(p_0) &= - p_0,
	\nonumber\\
	S(p_i) &= - e^{p_0/\kappa}p_i .
	\label{eq:bicross_antipode}
\end{align}

The Lorentz sector in the bicrossproduct basis closes as in the undeformed Poincar\'e algebra,
\begin{align}
	[M_i,M_j] &= i\epsilon_{ijk}M_k,
	\label{eq:bicross_rot_rot}
	\\
	[M_i,N_j] &= i\epsilon_{ijk}N_k,
	\label{eq:bicross_rot_boost}
	\\
	[N_i,N_j] &= -\,i\epsilon_{ijk}M_k,
	\label{eq:bicross_boost_boost}
\end{align}
and rotations act on momenta in the standard way,
\begin{align}
	[M_i,p_0] &= 0,
	\label{eq:bicross_rot_energy}
	\\
	[M_i,p_j] &= i\epsilon_{ijk}p_k.
	\label{eq:bicross_rot_momentum}
\end{align}
The deformation appears in the boost action on the translation sector:
\begin{align}
	[N_i,p_0] &= i p_i,
	\label{eq:bicross_boost_energy}
	\\
	[N_i,p_j] &= i\delta_{ij}
	\left[
	\frac{\kappa}{2}\left(1-e^{-2p_0/\kappa}\right)
	+\frac{\vec p^{\,2}}{2\kappa}
	\right]
	-\frac{i}{\kappa}p_i p_j.
	\label{eq:bicross_boost_momentum}
\end{align}
The boost coproduct is nonprimitive,
\begin{equation}\label{eq:bicross_boost_coproduct}
	\Delta(N_i)
	=
	N_i \otimes 1
	+
	e^{-p_0/\kappa} \otimes N_i
	+
	\frac{1}{\kappa}\epsilon_{ijk} p_j \otimes M_k ,
\end{equation}
whereas the rotation coproduct remains primitive,
\begin{equation}\label{eq:bicross_rotation_coproduct}
	\Delta(M_i)=M_i\otimes 1 + 1\otimes M_i.
\end{equation}
The corresponding bicrossproduct Casimir is
\begin{equation}\label{eq:bicross_casimir}
	{\cal C}_{\rm bicross}(p)
	=
	\left(2\kappa\sinh\frac{p_0}{2\kappa}\right)^2
	-
	e^{p_0/\kappa}\vec p^{\,2},
\end{equation}
so the free one-particle mass shell is ${\cal C}_{\rm bicross}(p)=m^2$.

This presentation is completely natural from the viewpoint of ordered plane waves: the variables $p_\mu$ are already the translation-generator eigenvalues, the deformed composition law \eqref{eq:deformed_addition} is built directly into the coalgebra, and the relation between ordered-plane-wave labels and bicrossproduct translation eigenvalues is one-to-one by construction.

\subsection{Classical basis}
\label{subsec:classical_basis}

In the present paper we instead work with the classical-basis translation generators, whose eigenvalues we denote by $P_\mu$ \cite{Kosinski:1994br, Borowiec:2009vb, Arzano:2022ewc}. These are related to the ordered-plane-wave labels $p_\mu$ by the nonlinear map
\begin{align}\label{eq:kappa_map}
	\begin{split}
		P_0 &= \kappa \sinh\!\left(\frac{p_0}{\kappa}\right)
		+ \frac{\vec{p}^{\,2}}{2\kappa} e^{p_0/\kappa}, 
		\\
		P_i &= p_i \, e^{p_0/\kappa}.
	\end{split}
\end{align}
It is also convenient to introduce the associated scalar
\begin{equation}\label{eq:P4_definition}
	P_4
	=
	\kappa \cosh\!\left(\frac{p_0}{\kappa}\right)
	-
	\frac{\vec p^{\,2}}{2\kappa} e^{p_0/\kappa},
	\qquad
	P_+ \equiv P_0 + P_4 = \kappa e^{p_0/\kappa}.
\end{equation}
The quantities $(P_0,\vec P,P_4)$ satisfy the de Sitter embedding relation
\begin{equation}\label{eq:dS_constraint}
	- P_0^2 + \vec P^{\,2} + P_4^2 = \kappa^2 .
\end{equation}
Accordingly, the same one-particle mass shell takes the quadratic classical-basis form
\begin{equation}\label{eq:classical_casimir}
	{\cal C}_{\rm cl}(P)=P_0^2-\vec P^{\,2}=P_4^2-\kappa^2,
\end{equation}
so that ${\cal C}_{\rm cl}(P)=m^2$. Thus $P_\mu$ may be viewed as classical-basis translation eigenvalues, while $p_\mu$ remain ordering-dependent auxiliary labels associated with the plane waves \eqref{eq:ordered_plane_waves}. At low energies $P_\mu\simeq p_\mu$, but away from that regime the relation is nonlinear and can become multivalued.

The ordered-plane-wave composition law \eqref{eq:deformed_addition} induces, via \eqref{eq:kappa_map} and \eqref{eq:P4_definition}, a nonabelian composition law for the classical-basis variables. Writing the first and second one-particle labels as $(P_\mu,P_+)$ and $(Q_\mu,Q_+)$, one finds \cite{Arzano:2022ewc}
\begin{align}
	(P \oplus Q)_0
	&=
	\frac{\kappa}{P_+}\,Q_0
	+
	\frac{\vec P\cdot \vec Q}{P_+}
	+
	\frac{1}{\kappa} P_0 Q_+,
	\label{eq:classical_addition_energy}
	\\
	(P \oplus Q)_i
	&=
	\frac{1}{\kappa} P_i Q_+ + Q_i,
	\label{eq:classical_addition_momentum}
\end{align}
together with
\begin{equation}\label{eq:classical_addition_Pplus}
	(P \oplus Q)_+ = \frac{1}{\kappa} P_+ Q_+ .
\end{equation}
These formulas define the translation coproduct in the classical basis. Equivalently, on tensor-product states,
\begin{align}
	\Delta(P_0)
	&=
	\frac{\kappa}{P_+}\otimes P_0
	+
	\sum_i \frac{1}{P_+} P_i \otimes P_i
	+
	\frac{1}{\kappa} P_0 \otimes P_+,
	\label{eq:coproduct_energy}
	\\
	\Delta(P_i)
	&=
	\frac{1}{\kappa} P_i \otimes P_+
	+
	1 \otimes P_i ,
	\label{eq:coproduct_momentum}
	\\
	\Delta(P_+)
	&=
	\frac{1}{\kappa} P_+ \otimes P_+ .
	\label{eq:coproduct_Pplus}
\end{align}
Here $P_+$ is not an additional independent translation generator; rather, it is the light-cone embedding coordinate $P_0+P_4$ needed to lift the projected four-vector $P_\mu$ to the full de Sitter momentum-space point within the chosen realization, and it packages the classical-basis coalgebra compactly. In a one-to-one regime $P_+$ is fixed by $P_\mu$, whereas in a multibranch regime the same four-vector $P_\mu$ can be accompanied by more than one real value of $P_+$.

The corresponding antipode, defined by the condition $P \oplus S(P)=0$, is
\begin{align}
	S(P_0) &= - P_0 + \frac{\vec P^{\,2}}{P_+},
	\nonumber\\
	S(P_i) &= - \frac{\kappa P_i}{P_+},
	\nonumber\\
	S(P_+) &= \frac{\kappa^2}{P_+}.
	\label{eq:classical_antipode}
\end{align}
In the undeformed limit one recovers the ordinary inverse four-momentum.

In the classical basis the Lorentz algebra acts linearly on $P_\mu$: the deformation is shifted from the one-particle Lorentz commutators to the coalgebra sector. In particular, the boost generators $N_i$ satisfy \cite{Kosinski:1994br, Borowiec:2009vb, Arzano:2022ewc}
\begin{align}
	[N_i, P_0] &= i P_i,
	\label{eq:boost_energy_action}
	\\
	[N_i, P_j] &= i \delta_{ij} P_0,
	\label{eq:boost_momentum_action}
\end{align}
while rotations act in the standard way. The deformation scale $\kappa$ remains observer independent. The boost coproduct is nevertheless nonprimitive:
\begin{equation}\label{eq:boost_coproduct}
	\Delta(N_i)
	=
	N_i \otimes 1
	+
	\frac{\kappa}{P_+}\otimes N_i
	+
	\sum_{j,k}\frac{1}{P_+}\epsilon_{ijk} P_j \otimes M_k ,
\end{equation}
where the rotation coproduct remains primitive,
\begin{equation}\label{eq:rotation_coproduct}
	\Delta(M_i)=M_i\otimes 1 + 1\otimes M_i.
\end{equation}
Thus the $\kappa$-Poincar\'e deformation survives in the multiparticle sector even though the one-particle Lorentz action on $P_\mu$ is the ordinary one.

\subsection{Comparison of the two bases and the Jacobian}
\label{subsec:comparison_jacobian}

The two presentations summarized above are related by the nonlinear change of variables \eqref{eq:kappa_map}. In the bicrossproduct basis the ordered-plane-wave labels are already the translation-generator eigenvalues, the deformed composition law is direct, and the mass shell is written in the nonlinear form \eqref{eq:bicross_casimir}. In the classical basis, by contrast, the same one-particle mass shell becomes quadratic as in \eqref{eq:classical_casimir}, while the nontriviality is shifted into the coalgebra through the $P_+$-dependent composition law.

As long as the map \eqref{eq:kappa_map} is locally invertible, these two descriptions are simply two equivalent local parametrizations of the same curved momentum space. A natural way to identify when this equivalence ceases to be one-to-one is through the Jacobian
\begin{equation}\label{eq:jacobian_definition}
	J_{\mu\nu}\equiv \frac{\partial P_\mu}{\partial p_\nu}.
\end{equation}
A direct computation gives (see Appendix~\ref{sec:jacobian_appendix} for the details of the computation)
\begin{equation}\label{eq:jacobian_determinant}
	\det\!\left(\frac{\partial P_\mu}{\partial p_\nu}\right)
	=
	e^{3p_0/\kappa}\,\frac{P_4}{\kappa}
	=
	\frac{P_+^3 P_4}{\kappa^4}.
\end{equation}
Therefore the change of variables \eqref{eq:kappa_map} is locally invertible whenever $\det J\neq 0$, and it ceases to be locally invertible precisely when
\begin{equation}\label{eq:jacobian_zero}
	\det J=0
	\qquad \Longleftrightarrow \qquad
	P_4=0.
\end{equation}
Using \eqref{eq:P4_definition}, this condition is equivalently
\begin{equation}\label{eq:critical_surface_p}
	\kappa \cosh\!\left(\frac{p_0}{\kappa}\right)
	-
	\frac{\vec p^{\,2}}{2\kappa}e^{p_0/\kappa}=0,
\end{equation}
or, after using $P_i=p_i e^{p_0/\kappa}$,
\begin{equation}\label{eq:critical_surface_P}
	e^{2p_0/\kappa}=\frac{\vec P^{\,2}}{\kappa^2}-1,
	\qquad\text{equivalently}\qquad
	P_+^2=\vec P^{\,2}-\kappa^2 .
\end{equation}
This will reappear in the next section as the condition for the fold of the inverse map.

In this sense, either basis provides a consistent description of the same curved momentum space as long as the Jacobian does not vanish.\footnote{A closely analogous redistribution of the deformation was emphasized in the nonrelativistic phase-space setting of Ref.~\cite{Nozari:2014qja}.  There, Darboux coordinates provide a local canonical chart in which the symplectic/Poisson algebra takes its standard form while the Hamiltonian is deformed, whereas in a noncanonical chart the Hamiltonian can retain a simple quadratic form at the price of a deformed Poisson structure.  The analogy with the present discussion is only structural: the bicrossproduct basis is similar in spirit to the Darboux description, in which the deformation is visible in the one-particle dispersion relation, while the classical basis is similar in spirit to a noncanonical description, in which the one-particle Casimir is quadratic but the nontriviality is shifted to the coproduct.  In the present relativistic setting, however, the relevant structure is not an ordinary symplectic chart transformation but a change of momentum-space parametrization within the $\kappa$-Poincar\'e Hopf algebra.} This is also the viewpoint emphasized by relative locality \cite{Amelino-Camelia:2011lvm,Amelino-Camelia:2011hjg}, where different momentum variables are regarded as different coordinate systems on the same momentum manifold rather than as a priori different one-particle theories. The mechanism studied in the rest of the paper begins precisely when the classical-basis parametrization \eqref{eq:kappa_map} ceases to be one-to-one. Since the corresponding coalgebra depends on $P_+=\kappa e^{p_0/\kappa}$ and therefore retains information about the chosen inverse branch, we shall work from now on in the classical basis.

\section{Coproduct-induced two-particle correlations}
\label{sec:coproduct_entanglement}

Having established the classical-basis presentation of the $\kappa$-Poincar\'e Hopf algebra, we now show how the nonprimitive coproduct gives rise to deformed two-particle momentum correlations once $\kappa$-effects are significant.

\subsection{Branch structure of the classical-basis map}
\label{subsec:branch_structure}

The main point established in subsection~\ref{subsec:comparison_jacobian} is that the classical-basis variables $P_\mu$ and the bicrossproduct variables $p_\mu$ are related by the nonlinear map \eqref{eq:kappa_map}, and that this map ceases to be locally invertible precisely when the Jacobian determinant \eqref{eq:jacobian_determinant} vanishes. Equivalently,
\begin{equation}\label{eq:branch_condition_from_jacobian}
	\det\!\left(\frac{\partial P_\mu}{\partial p_\nu}\right)=0
	\qquad \Longleftrightarrow \qquad
	P_4=0
	\qquad \Longleftrightarrow \qquad
	e^{2p_0/\kappa}=\frac{\vec P^{\,2}}{\kappa^2}-1.
\end{equation}
This is the fold at which the inverse relation $P_\mu=P_\mu(p)$ can develop more than one real local branch.

To make the branch structure explicit, we start again from the second line of \eqref{eq:kappa_map},
\begin{equation}\label{eq:p_from_P}
	p_i = P_i e^{-p_0/\kappa},
	\qquad
	\vec p^{\,2} = \vec P^{\,2} e^{-2p_0/\kappa}.
\end{equation}
Thus any possible multivaluedness of the inverse map $P\mapsto p$ is entirely governed by the $P_0$ relation in \eqref{eq:kappa_map}. Substituting \eqref{eq:p_from_P} into the first line of \eqref{eq:kappa_map} gives
\begin{equation}\label{eq:solve_fu}
	f\!\Big(\frac{p_0}{\kappa};\,\frac{|\vec P|}{\kappa}\Big)=\frac{P_0}{\kappa},
\end{equation}
where
\begin{equation}\label{eq:def_f}
	f\!\Big(\frac{p_0}{\kappa};\,\frac{|\vec P|}{\kappa}\Big)
	\equiv
	\sinh\!\Big(\frac{p_0}{\kappa}\Big)
	+
	\frac{\vec P^{\,2}}{2\kappa^2}\,e^{-p_0/\kappa}.
\end{equation}
Then inversion of \eqref{eq:kappa_map} for fixed $(P_0,\vec P)$ reduces to solving Eq.~\eqref{eq:solve_fu}. The number of real solutions is determined by the monotonicity of $f$. Differentiating with respect to $p_0$, we find
\begin{equation}\label{eq:crit_u}
	\frac{{\D}f}{{\D}p_0}=0
	\quad \Longrightarrow \quad
	e^{2p_0/\kappa}=\frac{\vec P^{\,2}}{\kappa^2}-1.
\end{equation}
As anticipated, this is the same condition already identified in \eqref{eq:branch_condition_from_jacobian} through the vanishing of the Jacobian determinant.

\begin{figure}[ht]
	\begin{subfigure}[b]{0.5\linewidth}
		\centering
		\includegraphics[width=1\linewidth]{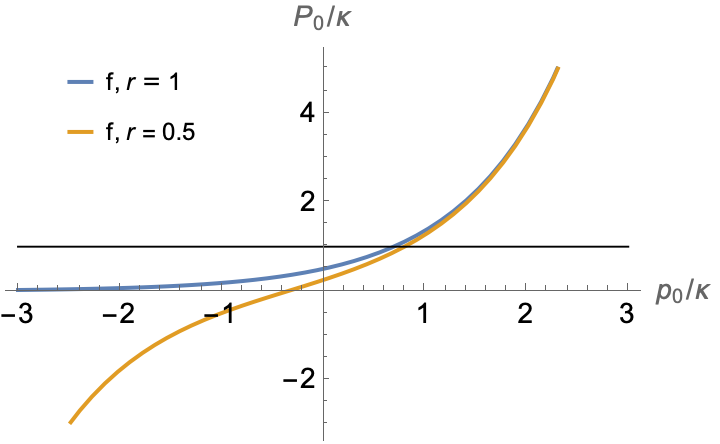}
		\vspace{2ex}
	\end{subfigure}
	\begin{subfigure}[b]{0.5\linewidth}
		\centering
		\includegraphics[width=1\linewidth]{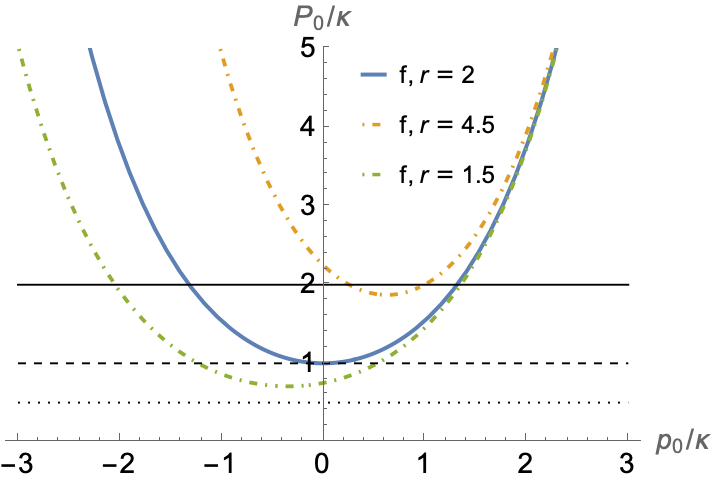}
		\vspace{2ex}
	\end{subfigure}
	\caption{The solid curves show $f\big(p_0/\kappa;r\big)$, defined in \eqref{eq:def_f}, plotted versus $p_0/\kappa$ for fixed $r\equiv|\vec P|/\kappa$ (equivalently fixed $r^2=|\vec P|^2/\kappa^2$). The horizontal lines indicate fixed values of $P_0/\kappa$, so intersections solve \eqref{eq:solve_fu} and correspond to real inverse branches $p_0(P_0,\vec P)$.}
	\label{fig:plot-f}
\end{figure}

Therefore:
\begin{itemize}
	\item If $\vec P^{\,2}< \kappa^2$, Eq.~\eqref{eq:crit_u} has no real solution, so ${{\D}f}/{{\D}p_0}$ never vanishes and $f$ is strictly increasing from $-\infty$ to $+\infty$. Hence \eqref{eq:solve_fu} has a unique real solution $p_0$ for each real $P_0$: the inverse map is single-valued (see the left panel of Fig.~\ref{fig:plot-f}). At the boundary $\vec P^{\,2}=\kappa^2$, one has $f=e^{p_0/\kappa}/2$, so there is a unique finite real solution only for $P_0>0$; $P_0\leq0$ lies outside the image of this coordinate patch.
	
	\item If $\vec P^{\,2}>\kappa^2$, Eq.~\eqref{eq:crit_u} has a real solution
	\begin{equation}\label{eq:u_star}
		p_{0*}=\frac{\kappa}{2}\ln\!\Big(\frac{\vec P^{\,2}}{\kappa^2}-1\Big),
	\end{equation}
	and
	\begin{equation}
		\kappa^2\frac{{\D}^2f}{{\D}p_0^2}\Big|_{p_0=p_{0*}}
		=
		\sqrt{\frac{\vec P^{\,2}}{\kappa^2}-1}>0,
	\end{equation}
	so $f$ has a minimum at $p_{0*}$. Consequently (see the right panel of Fig.~\ref{fig:plot-f}):
	\begin{itemize}
		\item for $\frac{P_0}{\kappa}>f\!\big(\frac{p_{0*}}{\kappa};\frac{|\vec P|}{\kappa}\big)$ there are {\it two real} solutions of \eqref{eq:solve_fu};
		
		\item for $\frac{P_0}{\kappa}=f\!\big(\frac{p_{0*}}{\kappa};\frac{|\vec P|}{\kappa}\big)$ there is one double real solution;
		
		\item for $\frac{P_0}{\kappa}<f\!\big(\frac{p_{0*}}{\kappa};\frac{|\vec P|}{\kappa}\big)$ there is no finite real solution.
	\end{itemize}
\end{itemize}

Equivalently, introducing $y\equiv e^{p_0/\kappa}>0$ turns the inversion problem into the quadratic equation
\begin{equation}\label{eq:y_quadratic}
	\kappa^2 y^2 - 2\kappa P_0 y + \bigl(\vec P^{\,2}-\kappa^2\bigr)=0.
\end{equation}
In the two-real-branch regime the two positive roots are
\begin{equation}\label{eq:Pplus_branches}
	y_\pm
	=
	\frac{P_0 \pm \sqrt{\kappa^2+P_0^2-\vec P^{\,2}}}{\kappa},
	\qquad
	P_+^{(\pm)}=\kappa y_\pm
	=
	P_0 \pm \sqrt{\kappa^2+P_0^2-\vec P^{\,2}}.
\end{equation}
The requirement $y_\pm>0$ is essential: for $\vec P^{\,2}>\kappa^2$ it is equivalent to $P_0>\sqrt{\vec P^{\,2}-\kappa^2}$, which is the same condition obtained from the minimum of $f$. Thus the same classical-basis four-vector $P_\mu$ can come with two distinct positive values of $P_+$ only in this region.

It is illuminating to note that the square root in \eqref{eq:Pplus_branches} is nothing but $|P_4|$: from the de Sitter relation \eqref{eq:dS_constraint} one has $P_4^2=\kappa^2+P_0^2-\vec P^{\,2}$, so that
\begin{equation}\label{eq:branch_P4}
	P_+^{(\pm)}=P_0\pm|P_4|=P_0+P_4^{(\pm)},
	\qquad
	P_4^{(\pm)}=\pm\sqrt{\kappa^2+P_0^2-\vec P^{\,2}}.
\end{equation}
The two branches are therefore the two signs of $P_4$, i.e. the two lifts of the same projected four-vector $P_\mu$ to the de Sitter hyperboloid \eqref{eq:dS_constraint} that lie in the patch $P_+>0$; they merge on the critical surface $P_4=0$ where the Jacobian \eqref{eq:jacobian_determinant} vanishes. In this sense, the multibranch structure is a projection degeneracy of the classical-basis coordinates: the one-particle label $P_\mu$ forgets the sign of $P_4$, while $P_+=P_0+P_4$ restores the missing sheet information. It is therefore natural, in the two-branch region, to use the sheet label
\begin{equation}\label{eq:sheet_label}
	\sigma=\operatorname{sgn}P_4=\pm1,
	\qquad
	P_4^{(\sigma)}=
	\sigma\sqrt{\kappa^2+P_0^2-\vec P^{\,2}},
	\qquad
	P_+^{(\sigma)}=P_0+P_4^{(\sigma)}.
\end{equation}
The two branches should not be regarded as additional particle species or independent internal quantum numbers; they are distinct lifts of the same projected four-momentum to the full momentum-space point $(P_\mu,P_4)$ within the chosen realization.

For a positive-energy on-shell one-particle state, $P_0=\sqrt{\vec P^{\,2}+m^2}$ and $P_0^2-\vec P^{\,2}=m^2$ fix the discriminant in \eqref{eq:Pplus_branches} to $\kappa^2+m^2$, and the two branches reduce to the compact form
\begin{equation}\label{eq:Pplus_onshell}
	P_+^{(\pm)}=\sqrt{\vec P^{\,2}+m^2}\,\pm\,\sqrt{\kappa^2+m^2},
\end{equation}
which becomes $P_+^{(\pm)}=|\vec P|\pm\kappa$ in the massless case. Since $\sqrt{\kappa^2+m^2}>0$, both roots are positive if and only if $|\vec P|>\kappa$, irrespective of the mass; equivalently, on the positive-energy mass shell the condition $P_0>\kappa\,f(p_{0*}/\kappa;|\vec P|/\kappa)=\sqrt{\vec P^{\,2}-\kappa^2}$ is automatically satisfied once $|\vec P|>\kappa$. Thus, for positive-energy on-shell states the onset of the two-positive-branch regime is governed by the single threshold $|\vec P|>\kappa$.

In Fig.~\ref{fig:plot-f} we illustrate the function $f$ defined in \eqref{eq:def_f} and the graphical solution of \eqref{eq:solve_fu} for fixed $(P_0,\vec P)$. For $|\vec P|^2<\kappa^2$ (left panel), $f$ is strictly increasing in $p_0$, so for any real $P_0$ the equation \eqref{eq:solve_fu} has a unique intersection and the inverse map $P\mapsto p$ is single-valued. At the boundary $|\vec P|^2=\kappa^2$, the range is instead $P_0>0$. For $|\vec P|^2>\kappa^2$ (right panel), $f$ develops a minimum at $p_{0*}$ given in \eqref{eq:u_star}, so \eqref{eq:solve_fu} can admit two real solutions: for $\frac{P_0}{\kappa}>f\big(\frac{p_{0*}}{\kappa};\frac{|\vec P|}{\kappa}\big)$ there are two intersections (two real branches), at equality there is a single double root, and for $\frac{P_0}{\kappa}<f\big(\frac{p_{0*}}{\kappa};\frac{|\vec P|}{\kappa}\big)$ there is no finite real solution. Moreover, when $\kappa^2<|\vec P|^2<2\kappa^2$ at least one of the two solutions has negative $p_0$ (blue/green dot-dashed examples), whereas for $|\vec P|^2>2\kappa^2$ both solutions can be positive for suitable choices of $P_0/\kappa$ (orange dot-dashed example).

Thus, for the specific map \eqref{eq:kappa_map} from ordered-plane-wave labels to classical-basis momenta, multibranch inversion with real auxiliary momenta can occur only if $|\vec P|>\kappa$ and $P_0$ lies above the minimum value attained by $f$. In the low-energy regime $|\vec P|\ll\kappa$ (and in particular as $\kappa\to\infty$) the inverse is single-valued and $P_\mu\simeq p_\mu$. This prepares the ground for the next subsection, where the $P_+$-dependent coproduct \eqref{eq:coproduct_energy}--\eqref{eq:coproduct_Pplus} turns this one-particle non-invertibility into branch-resolved two-particle correlations.

\subsection{From coproduct correlations to auxiliary branch expansions}
\label{subsec:correlated_states}

We now turn to the Hopf-algebraic input that ties the two subsystems together, namely the classical-basis coproduct
\eqref{eq:coproduct_energy}--\eqref{eq:coproduct_Pplus}. Because this coproduct depends explicitly on $P_+$, it is useful to distinguish first the branch-resolved one-particle states and only afterwards the coarse-grained states labeled by $P_\mu$ alone.

For a two-particle system with
$\mathcal{H}_2\equiv\mathcal{H}_1\otimes\mathcal{H}_1$, where $\mathcal{H}_1$ carries the one-particle
representation of the commuting operators $P_\mu$ and $P_+$, the deformed ``center-of-mass'' condition for $|\Psi\rangle\in\mathcal{H}_2$ is
\begin{equation}\label{eq:total_momentum_zero}
	\Delta(P_\mu)\,|\Psi\rangle = 0,
	\qquad |\Psi\rangle\in\mathcal{H}_2 .
\end{equation}
This replaces the undeformed constraint $P^{(1)}_\mu+P^{(2)}_\mu=0$.

\subsubsection{$P$-basis: branch-resolved correlated products}

Let $|P,k\rangle\in\mathcal{H}_1$ denote a simultaneous eigenstate of the one-particle operators $P_\mu$ and $P_+$,
\begin{equation}\label{eq:branch_resolved_P_state}
	P_\mu\,|P,k\rangle = P_\mu\,|P,k\rangle,
	\qquad
	P_+\,|P,k\rangle = P_+^{(k)}(P)\,|P,k\rangle,
\end{equation}
where $k$ labels the real inverse branches and $P_+^{(k)}(P)=\kappa e^{p_0^{(k)}(P)/\kappa}$. In a one-branch regime there is only one such label. In a multibranch regime, different $k$ correspond to the same coarse-grained $P_\mu$ but to different values of $P_+$.

Because $\Delta(P_\mu)$ is built from $P_\mu$ and $P_+$ and acts diagonally on branch-resolved product eigenstates, a sharp product eigenstate $|P^{(1)},k\rangle_1\otimes|P^{(2)},\ell\rangle_2$ satisfies \eqref{eq:total_momentum_zero} if and only if its eigenvalues obey the ordered classical-basis coproduct constraint. Since the coproduct is noncocommutative, this condition depends on the order of the tensor factors and here we use the ordering $|P^{(1)},k\rangle_1\otimes |P^{(2)},\ell\rangle_2$. Thus
\begin{align}
	\frac{\kappa}{P_+^{(1,k)}}\,P^{(2)}_0
	+
	\frac{\vec P^{\,(1)}\cdot \vec P^{\,(2)}}{P_+^{(1,k)}}
	+
	\frac{1}{\kappa} P^{(1)}_0 P^{(2,\ell)}_+
	&=
	0,
	\label{eq:energy_conservation}
	\\
	\frac{1}{\kappa}\,\vec{P}^{\,(1)} P^{(2,\ell)}_+
	+
	\vec{P}^{\,(2)}
	&=
	0,
	\label{eq:spatial_momentum_conservation}
\end{align}
equivalently, using the antipode \eqref{eq:classical_antipode},
\begin{align}
	P^{(2)}_0(k)
	&=
	-
	P^{(1)}_0
	+
	\frac{\vec P^{\,(1)2}}{P_+^{(1,k)}},
	\nonumber\\
	\vec P^{\,(2)}(k)
	&=
	-
	\frac{\kappa}{P_+^{(1,k)}}\,\vec P^{\,(1)},
	\nonumber\\
	P^{(2)}_+(k)
	&=
	\frac{\kappa^2}{P_+^{(1,k)}}.
	\label{eq:deformed_antipode}
\end{align}
Thus each resolved branch $k$ determines a $\kappa$-dependent back-to-back relation: in the undeformed limit
$\kappa\to\infty$ one recovers $P^{(2)}_0=-P^{(1)}_0$ and $\vec{P}^{\,(2)}=-\vec{P}^{\,(1)}$.

The sheet interpretation makes the branch dependence of the antipode transparent. Defining $P_4^{(a)}=P_+^{(a)}-P_0^{(a)}$ for each particle, the correlated partner determined by \eqref{eq:deformed_antipode} satisfies
\begin{align}
	P_4^{(2)}(k)
	&=
	P_+^{(2)}(k)-P_0^{(2)}(k)
	\nonumber\\
	&=
	\frac{\kappa^2}{P_+^{(1,k)}}
	+
	P_0^{(1)}
	-
	\frac{\vec P^{\,(1)2}}{P_+^{(1,k)}}
	\nonumber\\
	&=
	P_4^{(1,k)} .
	\label{eq:antipode_preserves_P4}
\end{align}
In the last step we used the de Sitter constraint and $P_+^{(1,k)}=P_0^{(1)}+P_4^{(1,k)}$. Note that here $P_4^{(2)}(k)$ denotes the $P_4$ value of the second particle selected by the branch $k$ of the first particle and it is not an independent branch label for the second particle. Thus the zero-total-momentum constraint does not pair a projected momentum with an arbitrary auxiliary branch: it selects the correlated partner on the corresponding lift of the projected momentum-space point. In this limited kinematical sense, the coproduct constraint is branch-resolving. For $\vec P^{\,(1)}\neq0$, the spatial part of \eqref{eq:deformed_antipode} also gives
\begin{equation}\label{eq:Pplus_tomography}
	P_+^{(1,k)}
	=
	-\kappa\,
	\frac{\vec P^{\,(1)2}}
	{\vec P^{\,(1)}\cdot\vec P^{\,(2)}} .
\end{equation}
Hence information invisible in the one-particle projected label $P_\mu$ can become visible in the two-particle correlation pattern fixed by the coproduct.

For a fixed sharp branch-resolved label $(P^{(1)}_\mu,P_+^{(1,k)})$, the constraint \eqref{eq:total_momentum_zero} fixes a unique correlated partner label $(P^{(2)}_\mu(k),P^{(2)}_+(k))$ through \eqref{eq:deformed_antipode}. Therefore, a branch-resolved constrained state is the product
\begin{equation}\label{eq:P_sharp_correlated_product}
	|\Psi_k\rangle
	=
	|P^{(1)},k\rangle_1\otimes\big|P^{(2)}(P^{(1)},k),\ell(k)\big\rangle_2 ,
\end{equation}
where $\ell(k)$ denotes the branch of the second particle selected by $P^{(2)}_+(k)$ when such a branch label is needed and in the sheet-label description this selection is equivalently characterized by \eqref{eq:antipode_preserves_P4}. One may of course superpose different branch-resolved or different coarse-grained values of $P^{(1)}$; this can produce ordinary momentum entanglement, exactly as in the undeformed case. For related discussions in deformed-symmetry settings, see also Refs.~\cite{Arzano:2026fbz,Lobo:2025vqt}.

\subsubsection{$p$-basis: auxiliary branch expansion}

As shown in subsection~\ref{subsec:branch_structure}, the nonlinear map \eqref{eq:kappa_map} need not be
one-to-one in certain regions: for a fixed coarse-grained translation-generator label $P_\mu$ there can exist multiple
real auxiliary labels $p_\mu^{(k)}$ and, equivalently, multiple values $P_+^{(k)}(P)$.
Accordingly, a chosen normalized one-particle state in the $P_\mu$ eigenspace can be written as
\begin{equation}\label{eq:P_expansion}
	|P\rangle = \sum_{k} \alpha_k \, |P,k\rangle
	= \sum_k \alpha_k \, |p_k\rangle,
	\qquad \sum_k|\alpha_k|^2=1,
\end{equation}
where each $|P,k\rangle$ is branch-resolved as in \eqref{eq:branch_resolved_P_state}, and each $|p_k\rangle$ is the unique auxiliary state corresponding to that same branch. The coefficients $\alpha_k$ are not fixed by the coproduct constraint or by multivaluedness alone; they characterize the particular state chosen within the degenerate one-particle eigenspace.

Applying the coproduct constraint branch by branch, the corresponding two-particle state becomes
\begin{equation}\label{eq:auxiliary_expanded_state}
	|\Psi\rangle
	=
	\sum_k \alpha_k\,
	|P^{(1)},k\rangle_1\otimes\big|P^{(2)}(P^{(1)},k),\ell(k)\big\rangle_2
	=
	\sum_k \alpha_k\,
	|p_k\rangle_1\otimes|\tilde p_k\rangle_2 ,
\end{equation}
where $P_\mu(\tilde p_k)=P^{(2)}_\mu(P^{(1)},k)$ and $P_+(\tilde p_k)=P^{(2)}_+(k)$. This makes explicit that, once the branch structure of \eqref{eq:kappa_map} is present, a state sharp only in the coarse-grained label $P_\mu$ becomes a branch-correlated two-particle state.

To make this structure explicit, consider the two-real-branch regime, in which
\eqref{eq:P_expansion} reduces to
\begin{equation}\label{eq:P_two_branch_expansion}
	|P\rangle=\alpha_+|p^{(+)}\rangle+\alpha_-|p^{(-)}\rangle,
	\qquad
	|\alpha_+|^2+|\alpha_-|^2=1,
\end{equation}
with
\begin{equation}\label{eq:two_branches}
	p_i^{(\pm)}=P_i\, e^{-p_0^{(\pm)}/\kappa},
	\qquad
	P_+^{(\pm)}=\kappa e^{p_0^{(\pm)}/\kappa}.
\end{equation}
In this case the constrained two-particle state takes the form
\begin{equation}\label{eq:auxiliary_expanded_state_two_branch}
	|\Psi\rangle
	=
	\alpha_+\,|p^{(+)}\rangle_1\otimes|\tilde p^{(+)}\rangle_2
	+
	\alpha_-\,|p^{(-)}\rangle_1\otimes|\tilde p^{(-)}\rangle_2,
\end{equation}
with $P_\mu(\tilde p^{(\pm)})=P^{(2)}_\mu(P^{(1)},\pm)$ determined by \eqref{eq:deformed_antipode}. In the high-momentum regime where the inverse map becomes multi-valued, a state sharp only in the coarse-grained classical-basis momentum label therefore acquires a nontrivial auxiliary branch expansion.

In this way, the purely kinematical content of the construction is twofold. In the branch-resolved classical basis, the coproduct fixes the deformed correlation between a given branch and its correlated partner. In the coarse-grained description, the same $P_\mu$ label can support several such branch-resolved products, producing the nontrivial expansion \eqref{eq:auxiliary_expanded_state}. We stress that the coproduct first gives a $\kappa$-deformed correlated product structure: for a sharp branch-resolved input the constrained state \eqref{eq:P_sharp_correlated_product} is separable. A coherent superposition over distinct constrained configurations, such as \eqref{eq:auxiliary_expanded_state_two_branch} with both coefficients nonzero, can of course be entangled in the usual bipartite sense. The deformation fixes the modified back-to-back relation \eqref{eq:deformed_antipode} and the branch dependence of the allowed components; it does not by itself determine the amplitudes $\alpha_k$. This raises the natural question of how the variables $P_\mu$ and $p_\mu$ should be interpreted operationally, to which we now turn.

It is worth noting that, since the deformation scale $\kappa$ is invariant under the $\kappa$-deformed Lorentz sector, the coproduct-induced two-particle momentum correlations found above are preserved under deformed boosts and are therefore not an observer-dependent artifact. In Appendix~\ref{sec:lorentz_covariance} we prove this explicitly: using only the Hopf-algebraic structure of the classical basis of $\kappa$-Poincar\'e symmetry, we give a fully algebraic demonstration that the corresponding coproduct-induced two-particle states transform covariantly under $\kappa$-deformed Lorentz transformations.

\subsection{Physical interpretation and observables}
\label{subsec:operational_prospects}

In the discussion above we work with the {\it real} auxiliary solutions $p_k$ of the inverse problem
$P_\mu=P_\mu(p)$ (setting aside complex branches when they arise). Real solutions are singled out because
they admit, in principle, a standard quantum-mechanical interpretation in terms of self-adjoint operators
$p_\mu$ and thus a potentially meaningful branch label. The preceding discussion suggests a conservative interpretation of the branch-sensitive variable: $P_+$ is not an additional independent translation generator, but the light-cone embedding coordinate $P_0+P_4$ needed to lift the projected classical-basis four-momentum $P_\mu$ to the full de Sitter momentum-space point. When the projection $(P_\mu,P_4)\mapsto P_\mu$ is one-to-one, this lift is fixed by $P_\mu$; when it is two-to-one, the same $P_\mu$ admits two lifts distinguished by $\operatorname{sgn}P_4$, and the coproduct is sensitive to this otherwise hidden sheet data.

It is also worth keeping in mind the
relative-locality viewpoint \cite{Amelino-Camelia:2011hjg}, according to which different momentum parametrizations of the same
curved momentum space should not be assigned an absolute meaning independently of the full operational setup.
From that perspective, statements formulated directly in terms of $P_\mu$ or $p_\mu$ should ultimately be assessed
through the momentum-space geometry and through the measurement framework used to probe it. In the present paper
we do not need to decide this issue in full generality. Rather, our aim is more limited: to identify, within the
classical-basis coproduct construction developed above, what the multibranch regime implies for two-particle states,
and to spell out how that structure may be read under different operational assumptions. With this in mind, the discussion naturally leads to two logically consistent viewpoints on the status of the variables $(P_\mu,p_\mu)$ and, consequently, on the meaning of the branch expansions discussed above:
\begin{enumerate}
	\renewcommand{\theenumi}{\Alph{enumi}}
	\renewcommand{\labelenumi}{\theenumi.}
	
	\item {\bf $P_\mu$ observable, $p_\mu$ auxiliary.}
	The directly accessible one-particle labels are the classical-basis translation-generator eigenvalues $P_\mu$. At low energies $P_\mu\simeq p_\mu$ and, in the one-to-one regime, the inverse map is single-valued, so the state is just the branch-resolved correlated product \eqref{eq:P_sharp_correlated_product}. In a multivalued regime the same projected label $P_\mu$ may correspond to two different lifts $(P_\mu,P_4^{(\sigma)})$. Since the classical-basis coproduct depends on $P_+=P_0+P_4$, it resolves this lift data in the composite sector: a state sharp only in the projected label $P_\mu$ can still take the branch-correlated form \eqref{eq:auxiliary_expanded_state}. In this viewpoint, the directly meaningful content remains the resulting deformed pattern of two-particle momentum correlations.
	
	\item {\bf $p_\mu$ observable, $P_\mu=P_\mu(p)$ derived.}
	Alternatively, detectors may couple directly to $p_\mu$, so $p_\mu$ are self-adjoint observables and
	$P_\mu\equiv P_\mu(p)$ are well-defined derived quantities via \eqref{eq:kappa_map}. Then
	multi-branching implies that distinct measured values $p_k$ can correspond to the same projected classical-basis eigenvalue $P_\mu$. In this case the expansions \eqref{eq:P_expansion} and \eqref{eq:auxiliary_expanded_state} are operationally meaningful as decompositions into different auxiliary branches associated with the same fixed $P_\mu$.
\end{enumerate}

Independently of the chosen viewpoint, the origin of the two-particle correlations is purely kinematical:\footnote{For a distinct kinematical effect associated with a quantum-group deformation of rotational symmetry, see Ref.~\cite{Arzano:2026fbz}. For entangled states generated by a different, dynamical open-system mechanism that can occur already at small $|\vec P|$ and does not rely on multi-branch inversion, see Ref.~\cite{Lobo:2025vqt}. This differs from the purely kinematical mechanism discussed here, which concerns coproduct-induced correlations together with the high-momentum regime $|\vec P|>\kappa$ where real multi-branch inversion can occur.} the Hopf-algebraic coproduct fixes the action of translations on ${\cal H}_2$ and hence correlates the two subsystems, while multi-branching of the map \eqref{eq:kappa_map} allows the same coarse-grained state to be resolved into several branch-resolved components. Accordingly, the branch-expanded description requires (i) at least two {\it real} inverse branches of \eqref{eq:kappa_map} for the relevant label, equivalently two positive values of $P_+$ for the same $P_\mu$ (for the parametrization used here: $|\vec P|>\kappa$ and $P_0/\kappa>f(p_{0*}/\kappa,|\vec P|/\kappa)$, or equivalently $P_0>\sqrt{\vec P^{\,2}-\kappa^2}$, with $p_{0*}$ given by \eqref{eq:u_star}), and (ii) a measurement framework in which the auxiliary labels or their branch-sensitive consequences are meaningful. Since (i) depends on the global invertibility of the chosen parametrization $P_\mu(p)$, changing the auxiliary realization can shift (or remove) the region where real multi-branching occurs, even though the Hopf-algebraic structures in $P_\mu$ are unchanged. In this sense, the present mechanism is best viewed as a realization-dependent kinematical feature of the classical-basis description. The relative-locality perspective remains an important guide in interpreting such realization dependence, but the specific question addressed here is the fate of the two-particle quantum-state construction once the map \eqref{eq:kappa_map} ceases to be one-to-one.

It is tempting to ask whether this branch-expanded correlated structure could be accessed in a controlled setting.
The relevant two-particle states obey the deformed ``center-of-mass'' constraint $\Delta(P_\mu)|\Psi\rangle=0$
in \eqref{eq:total_momentum_zero}, together with a regime in which the realization \eqref{eq:kappa_map} admits
two {\it real} inverse branches. This is intrinsically a high-momentum effect: for $|\vec P|\ll\kappa$ (or
$\kappa\to\infty$) the inverse is single-valued and $P_\mu\simeq p_\mu$, so the branch-expanded description collapses to a single term. As an order-of-magnitude guide, current collider center-of-mass energies are of order $\sqrt{s}\sim 10^{4}\,\mathrm{GeV}$, so any {\it unsuppressed} $\kappa$-effects in standard observables would suggest $\kappa \gtrsim 10^{4}\,\mathrm{GeV}$,
though this is not a universal bound and depends on how the deformation enters the measured sector; in
particular, the present mechanism requires access to the regime where real multi-branch inversion occurs
for \eqref{eq:kappa_map}.\footnote{While laboratories are conceptually better suited for controlled correlation measurements, some of the strongest bounds on $\kappa$-deformed kinematics come from astrophysical time-of-flight and threshold analyses: ultra-high-energy cosmic rays reach $\sim 10^{11}\,\mathrm{GeV}$ \cite{Amelino-Camelia:1997ieq,Amelino-Camelia:2000cpa,Jacobson:2001tu,Jacobson:2005bg,Liberati:2009pf,Liberati:2013xla}. Such constraints are often phrased in terms of basis-dependent modified dispersion relations or related threshold effects. In the present classical basis the Lorentz Casimir is the standard quadratic expression $P_0^2-\vec P^{\,2}=P_4^2-\kappa^2$, while the same momentum-space geometry is often discussed in bicrossproduct coordinates, where the Casimir takes a nonlinear form \cite{Kosinski:1994br,Arzano:2022ewc}. We do not pursue a detailed comparison with these bounds here.}

To make these prospects quantitative, it is useful to separate the two regimes by the relevant fractional $\kappa$-correction.

\noindent\textit{(i) Single-branch regime, $|\vec P|<\kappa$.} Here the inverse map is one-to-one and the only effect is the $\kappa$-deformation of the back-to-back relation \eqref{eq:deformed_antipode}. The partner spatial momentum has magnitude $|\vec P^{\,(2)}|=(\kappa/P_+^{(1)})\,|\vec P^{\,(1)}|=e^{-p_0^{(1)}/\kappa}\,|\vec P^{\,(1)}|$, so the leading departure from the undeformed back-to-back configuration $\vec P^{\,(2)}=-\vec P^{\,(1)}$ is
\begin{equation}\label{eq:lowE_deviation}
	\frac{|\vec P^{\,(2)}|}{|\vec P^{\,(1)}|}
	=
	e^{-p_0^{(1)}/\kappa}
	=
	1-\frac{p_0^{(1)}}{\kappa}+\mathcal{O}\!\big(\kappa^{-2}\big),
\end{equation}
i.e. a fractional correction of order $E/\kappa$, the familiar linear-in-energy DSR suppression. For $E\sim 1\,\mathrm{TeV}$ and $\kappa\sim\Mpl\simeq1.2\times10^{19}\,\mathrm{GeV}$ this is $\sim10^{-16}$, while for the highest-energy cosmic rays, $E\sim10^{11}\,\mathrm{GeV}$, it is still only $\sim10^{-8}$. Such corrections are in principle present at any energy, but are unobservably small unless $\kappa$ lies far below $\Mpl$.

\noindent\textit{(ii) Two-branch regime, $|\vec P|>\kappa$.} This is the genuinely new window opened by the multivaluedness of \eqref{eq:kappa_map}, and it is intrinsically trans-$\kappa$: it requires individual momenta above the deformation scale. If $\kappa\sim\Mpl$, this is super-Planckian and inaccessible; even ultra-high-energy cosmic rays fall short by some eight orders of magnitude, so the effect is relevant only in scenarios with a lowered scale $\kappa\ll\Mpl$. When the regime is reached, the two branches \eqref{eq:Pplus_onshell} are genuinely separated. For an on-shell massless particle the corresponding auxiliary energies differ by
\begin{equation}\label{eq:branch_separation}
	\Delta p_0\equiv p_0^{(+)}-p_0^{(-)}
	=
	\kappa\,\ln\!\frac{P_+^{(+)}}{P_+^{(-)}}
	=
	\kappa\,\ln\!\frac{|\vec P|+\kappa}{|\vec P|-\kappa},
\end{equation}
which diverges as $|\vec P|\to\kappa^+$ (because the lower branch approaches the boundary $P_+=0$) and decreases as $2\kappa^2/|\vec P|$ deep in the trans-$\kappa$ regime. The associated branch-dependent partners carry momenta whose magnitudes differ by the $\mathcal{O}(1)$ ratio
\begin{equation}\label{eq:partner_ratio}
	\frac{|\vec P^{\,(2,+)}|}{|\vec P^{\,(2,-)}|}
	=
	\frac{P_+^{(-)}}{P_+^{(+)}}
	=
	\frac{|\vec P|-\kappa}{|\vec P|+\kappa}.
\end{equation}
Thus, in this regime, the two branches predict markedly different correlated partners for the same coarse-grained $P_\mu$. It is this $\mathcal{O}(1)$ branch separation near $|\vec P|\sim\kappa$, rather than the $\mathcal{O}(E/\kappa)$ correction of regime (i), that is the characteristic signature of the multibranch structure.

In summary, the mechanism discussed here is entirely kinematical: the classical-basis $\kappa$-Poincar\'e coproduct
\eqref{eq:coproduct_energy}--\eqref{eq:coproduct_Pplus} fixes the translation action on ${\cal H}_2$
and enforces branch-resolved two-particle momentum correlations through
\eqref{eq:energy_conservation}--\eqref{eq:deformed_antipode}, while a multi-valued inverse of the map
\eqref{eq:kappa_map} (when it admits multiple {\it real} branches) allows the same coarse-grained state to
be rewritten as a nontrivial auxiliary branch expansion. From a phenomenological viewpoint, any laboratory
``setup'' should be regarded only as a cautious {\it gedanken} guide: in principle, a collider-type
environment is the most natural arena, since it permits repeatable state preparation and controlled
correlation measurements.

\section{Relation to minimal length scenarios}
\label{sec:GUP_relation}

In this section we show that the same purely kinematical correlation mechanism found in
Sec.~\ref{sec:coproduct_entanglement} does not arise in the standard regular self-adjoint
nonrelativistic minimal-length models considered below.

A particularly transparent comparison with our discussion arises in the nonrelativistic
setting in which only the spatial sector is deformed and energy does not mix with momentum. A common
starting point is a modified Heisenberg algebra written directly in terms of the translation-generator
operator $\hat P$,
\begin{equation}\label{eq:NR_GUP_general}
	[\hat x,\hat P] = i\hbar\,G(\hat P),
\end{equation}
where $G$ is a model-dependent function. To make contact with an auxiliary description one introduces a canonical momentum $\hat p$ satisfying $[\hat x,\hat p]=i\hbar$, and
represents the momentum operator as a function of $\hat p$,
\begin{equation}\label{eq:P_as_function_of_p}
	\hat P = F(\hat p).
\end{equation}
Using the standard identity for a canonical pair, $[\hat x,F(\hat p)]=i\hbar\,F'(\hat p)$, the
commutator \eqref{eq:NR_GUP_general} implies the operator functional relation
$F'(\hat p)=G\!(F(\hat p))$, i.e. at the level of the associated classical map,
\begin{equation}\label{eq:NR_dPdp_general}
	\frac{{\D}P}{{\D}p}=G(P).
\end{equation}
Thus specifying $G(P)$ determines the momentum map $P=F(p)$ up to an integration constant fixed by the
low-momentum matching $P\simeq p$. At this stage, it is tempting to expect that a nonlinear map
$P=F(p)$ might generically lead to multi-branch inversion (and therefore, in suitable composite-system
setups, to branch-expanded descriptions analogous to those discussed in
Sec.~\ref{sec:coproduct_entanglement}). However, whether such a genuinely local, non-periodic two-branch structure can occur in a consistent self-adjoint realization is a subtle question, because it depends not only on the algebraic form of $G$ but also on the admissible spectral/domain restrictions needed to keep $\hat P$ self-adjoint and the commutator \eqref{eq:NR_GUP_general} regular. Below we
review standard minimal-length examples and then show that, under the natural regularity/positivity requirements on $G(\hat P)$, these nonrelativistic models do not support the kind of smooth, pole-free, locally two-real-branch inversion that appears in our $\kappa$-Poincar\'e setting.

\subsection{Polymer quantum mechanics}

A closely related example is provided by polymer quantum mechanics, which fits into the general
nonrelativistic framework \eqref{eq:NR_GUP_general} with the choice \cite{Ashtekar:2002sn,Corichi:2007tf,Corichi:2012bg,Gorji:2015kta}
\begin{equation}\label{eq:polymer_G_of_P}
	G(\hat P)=\sqrt{1-\lambda^{2}\hat P^{\,2}},
\end{equation}
where $\lambda$ sets the polymer scale. In a self-adjoint realization one restricts to the real domain
of \eqref{eq:polymer_G_of_P}, so the spectrum of $\hat P$ is bounded by $|P|\le 1/\lambda$.

Substituting \eqref{eq:polymer_G_of_P} in \eqref{eq:NR_dPdp_general} and fixing the integration constant
by the low-momentum matching $P\simeq p$ gives the exact all-orders momentum map
\begin{equation}\label{eq:polymer_P_of_p_sin}
	P = \frac{1}{\lambda}\sin\!\big(\lambda p\big).
\end{equation}
The low-momentum expansion of \eqref{eq:polymer_P_of_p_sin} reads
$P= p - \lambda^{2}p^{3}/6+\mathcal{O}(\lambda^{4}p^{5})$. One typically restricts to a single chart (an open
monotonicity interval) so that the inverse relation $p=p(P)$ becomes single-valued. Concretely, choosing one monotonic branch of $\sin$ corresponds to taking, e.g. $p \in \left(-\pi/(2\lambda),\pi/(2\lambda)\right)$. On such a chart the periodic branch label disappears. In this
restricted description both $p$ and $P$ remain bounded: $p$ is bounded by construction, and
$P=(1/\lambda)\sin(\lambda p)$ stays within $|P|\le 1/\lambda$.

The multi-valued structure of the inverse relation $p=p(P)$ follows directly from
\eqref{eq:polymer_P_of_p_sin}: for fixed real $P$ with $|P|\le 1/\lambda$,
\begin{equation}\label{eq:polymer_inverse_branches}
	p(P)=\frac{1}{\lambda}\arcsin(\lambda P)+\frac{2\pi n}{\lambda},
	\qquad
	p(P)=\frac{1}{\lambda}\big(\pi-\arcsin(\lambda P)\big)+\frac{2\pi n}{\lambda},
	\qquad n\in\mathbb{Z},
\end{equation}
reflecting the periodicity of $P=(1/\lambda)\sin(\lambda p)$ and the resulting global identification
of auxiliary labels.

\subsection{Generalized uncertainty principle}

Another standard class of minimal-length models is provided by generalized uncertainty principle (GUP)
deformations, i.e. different prescriptions for $G(\hat P)$ in \eqref{eq:NR_GUP_general}.

A widely used perturbative ansatz is \cite{Kempf:1994su,Kempf:1996nk}
\begin{equation}\label{eq:GUP_commutator}
	G(\hat P)=1+\beta \hat P^{\,2},
\end{equation}
with $\beta>0$. Substituting \eqref{eq:GUP_commutator} in \eqref{eq:NR_dPdp_general} and integrating gives
\begin{equation}\label{eq:P_of_p_tan}
	\hat P = \frac{1}{\sqrt{\beta}}\tan\!\big(\sqrt{\beta}\hat p\big),
\end{equation}
whose low-momentum expansion reads
\begin{equation}\label{eq:GUP_tan_expansion}
	P = p + \frac{\beta}{3}\,p^{3}+\mathcal{O}(\beta^{2}p^{5}).
\end{equation}
The inverse relation is multi-valued at the classical level purely due to the periodicity of $\tan$,
\begin{equation}\label{eq:tan_inverse_branches}
	p(P) = \frac{1}{\sqrt{\beta}}\arctan(\sqrt{\beta}\,P)+\frac{n\pi}{\sqrt{\beta}},
	\qquad n\in\mathbb{Z}.
\end{equation}
As in the polymer case, one may restrict to a single chart for the argument of $\tan$, $p \in (-\pi/(2\sqrt{\beta}),\pi/(2\sqrt{\beta}))$, so that the map becomes one-to-one and the branch label disappears. In this restricted description $p$ is bounded, while $|P|\to\infty$ as $p\to\pm \pi/(2\sqrt{\beta})$.

The multivalued inverse maps \eqref{eq:polymer_inverse_branches} (polymer) and
\eqref{eq:tan_inverse_branches} (GUP) arise from the global non-injectivity of periodic functions
(such as $\sin$ and $\tan$): a single momentum label corresponds to infinitely many preimages differing by
integer multiples of the period (or by shifts between branches). This ``global periodic'' multivaluedness
is qualitatively different from the present $\kappa$-Poincar\'e map \eqref{eq:kappa_map}, where the two real solutions
of \eqref{eq:solve_fu} appear because the function $f$ develops a genuine local minimum, producing a
finite two-branch structure in a connected high-momentum region.

To make this distinction explicit, it is useful to consider non-perturbative deformations that share the
same formal low-momentum series expansion but differ in their global structure. For example, one may take \cite{Pedram:2011gw,Pedram:2012my}
\begin{equation}\label{eq:GUP_commutator_np}
	G(\hat P)=\frac{1}{1-\beta \hat P^{\,2}},
\end{equation}
which admits the formal expansion
$(1-\beta \hat P^{2})^{-1}=1+\beta \hat P^{2}+\beta^{2}\hat P^{4}+\cdots$ on a suitable low-momentum
domain. Substituting \eqref{eq:GUP_commutator_np} into \eqref{eq:NR_dPdp_general} and integrating gives
(fixing $P\simeq p$ near $P=0$)
\begin{equation}\label{eq:GUP_np_p_of_P}
	p(P) = P - \frac{\beta}{3}\,P^{3}.
\end{equation}
Formally, \eqref{eq:GUP_np_p_of_P} has turning points at $P_*=\pm 1/\sqrt{\beta}$, with extrema
\begin{equation}\label{eq:GUP_np_extrema}
	p_{\max}=\frac{2}{3\sqrt{\beta}},
	\qquad
	p_{\min}=-\frac{2}{3\sqrt{\beta}}.
\end{equation}
Equivalently, for fixed $p$ the inversion problem is a depressed cubic for $P$ with three real algebraic
roots for $|p|<2/(3\sqrt{\beta})$. However, the turning points coincide with the pole of
$G(\hat P)$ in \eqref{eq:GUP_commutator_np}. A consistent self-adjoint realization therefore requires
restricting the spectrum of $\hat P$ away from the singularity; in the most conservative implementation
one imposes
\begin{equation}\label{eq:GUP_np_subcritical}
	|P|<\frac{1}{\sqrt{\beta}}.
\end{equation}
On this admissible domain one has ${\D}p/{\D}P=1-\beta P^{2}>0$, so \eqref{eq:GUP_np_p_of_P} is strictly
monotonic and the physical inverse $P=P(p)$ is single-valued: although the algebraic cubic admits
multiple real roots, only the root lying in \eqref{eq:GUP_np_subcritical} is compatible with a regular
realization of \eqref{eq:GUP_commutator_np}.

Finally, let us comment on Ref.~\cite{Moradpour:2026gib}, which motivated part of the discussion here. That
work analyzes a cubic truncation of the momentum map (effectively of the form $P\simeq p(1+\beta p^{2})$)
and emphasizes that the inverse problem becomes a third-order equation with two complex roots, which is
then interpreted as a novel feature and linked to non-Hermitian operators. From the viewpoint adopted in
this paper, complex branches primarily signal that one has either (i) truncated an underlying all-orders
map in a way that changes its global properties, or (ii) moved outside the real domain of a chosen
parametrization. If one insists on standard quantum-mechanical observables represented by self-adjoint
operators, then treating $p$ as observable requires restricting to real spectra and to a
domain/functional calculus where $\hat P=F(\hat p)$ is self-adjoint; in this sense, complex
``eigenvalues'' are not automatically endowed with an operational meaning.

\subsection{$\kappa$-Poincar\'e vs. nonrelativistic}

The nonrelativistic framework \eqref{eq:NR_GUP_general}--\eqref{eq:NR_dPdp_general} makes a key point
transparent: in a regular self-adjoint realization one typically requires $G(\hat P)$ to be a positive
operator on the physical domain (so that the commutator is well-defined and does not change sign),
\begin{equation}\label{eq:G_positive_domain}
	G(\hat P)>0 \quad \text{on the spectrum of }\hat P .
\end{equation}
At the level of the associated classical map, \eqref{eq:G_positive_domain} means $G(P)>0$ throughout the
admissible real domain of $P$. But then \eqref{eq:NR_dPdp_general} immediately implies strict monotonicity:
\begin{equation}\label{eq:monotonicity}
	\frac{{\D}P}{{\D}p}=G(P)>0
	\quad\Rightarrow\quad
	P(p)\ \text{is strictly increasing on the admissible domain.}
\end{equation}
A strictly monotone map is one-to-one, hence it admits a single-valued inverse $p=p(P)$ on that domain.
Equivalently, rewriting \eqref{eq:NR_dPdp_general} as ${\D}p/{\D}P=1/G(P)$ shows that ${\D}p/{\D}P>0$ wherever $G(P)>0$,
so $p(P)$ is strictly increasing and cannot develop folds. Therefore, under the standard regularity and
positivity requirement \eqref{eq:G_positive_domain}, {\it genuine local two-real-branch inversion is impossible} in the nonrelativistic setting: one cannot have two distinct real values of $p$ mapping to the same admissible real value of $P$ without leaving the domain where $G(\hat P)$ stays positive and
regular.

This observation clarifies the status of the examples above. Polymer quantum mechanics and the exact
$\tan$-completion of the perturbative GUP do exhibit multi-valued inverses, but these are global/periodic. The non-perturbative model \eqref{eq:GUP_commutator_np} develops turning points only at the
price of a pole in $G(\hat P)$; enforcing self-adjointness and regularity by restricting the spectrum
removes any physically admissible two-real-branch regime.

By contrast, our $\kappa$-Poincar\'e mechanism is genuinely different (and,
in this sense, stronger): it is tied to the specific nonlinear relativistic
map \eqref{eq:kappa_map} between the classical-basis translation generators
$\hat P_\mu$ and the ordered-plane-wave labels $p_\mu$. The possibility of
two real inverse branches for fixed classical-basis momentum label $P_\mu$
is controlled by the nontrivial dependence of $P_0$ on both $p_0$ and
$\vec p^{\,2}$ in \eqref{eq:kappa_map}, which leads to the non-monotonic
behavior of the function $f$ in \eqref{eq:def_f} and hence to two real
solutions of \eqref{eq:solve_fu} in a high-momentum regime. Importantly,
this happens without a pole or divergence in the map \eqref{eq:kappa_map}
itself: the map remains smooth and finite at the fold, while its Jacobian
loses rank at $P_4=0$. This is different from the nonrelativistic singular
examples above, where the would-be turning point is tied to a pole of the
commutator function $G(P)$, and it also does not reduce to a
one-dimensional flow of the type \eqref{eq:NR_dPdp_general}.

Operationally, this difference matters. In the nonrelativistic setting, once one enforces
\eqref{eq:G_positive_domain} the auxiliary map is one-to-one on the physical domain, so rewriting
constraints in an auxiliary $p$-basis does not generically produce the kind of branch-expanded
descriptions that arise in our relativistic setting. In the $\kappa$-Poincar\'e setting,
instead, the smooth real two-branch inverse map in the high-momentum regime makes such branch-expanded
descriptions possible already kinematically, while the classical-basis coproduct simultaneously enforces the deformed
two-particle correlations of Sec.~\ref{sec:coproduct_entanglement}. In this sense, the mechanism discussed
there is genuinely specific to the relativistic $\kappa$-Poincar\'e framework.

\section{Summary and conclusion}\label{sec:summary}

We have analyzed a kinematical source of two-particle momentum correlations in $\kappa$-Minkowski spacetime. The starting point was the fact that the same curved momentum space admits different useful parametrizations. In the bicrossproduct basis the ordered-plane-wave labels $p_\mu$ coincide with the translation-generator eigenvalues, and the corresponding description is one-to-one. In the classical basis the translation eigenvalues are instead the variables $P_\mu$, related to $p_\mu$ by the nonlinear map \eqref{eq:kappa_map}. The one-particle Lorentz action on $P_\mu$ is then ordinary, but the deformation reappears in the multiparticle sector through the nonprimitive coproduct \eqref{eq:coproduct_energy}--\eqref{eq:coproduct_Pplus}.

Imposing the deformed ``center-of-mass'' constraint $\Delta(P_\mu)|\Psi\rangle=0$ fixes a $\kappa$-dependent back-to-back relation between the two particles. For a branch-resolved one-particle label $(P_\mu,P_+)$ the partner is determined by the antipode, Eq.~\eqref{eq:deformed_antipode}, so the constrained state is a deformed correlated product. This part of the construction is already present in the single-branch regime and reduces smoothly to the usual opposite-momentum relation as $\kappa\to\infty$.

The additional structure appears when the classical-basis map is not globally one-to-one. The inverse problem is governed by the function \eqref{eq:def_f}, or equivalently by the quadratic equation \eqref{eq:y_quadratic} for $y=e^{p_0/\kappa}$. In the patch $P_+>0$, two real auxiliary branches exist when $|\vec P|>\kappa$ and $P_0>\sqrt{\vec P^{\,2}-\kappa^2}$. For positive-energy on-shell states this reduces to the simple threshold $|\vec P|>\kappa$. Geometrically, this is a projection degeneracy: the same projected classical-basis label $P_\mu$ can be accompanied by two distinct positive values of $P_+=P_0+P_4$, corresponding to the two signs of the de Sitter embedding coordinate $P_4$. The coproduct distinguishes these two lifts, and the antipode preserves the corresponding $P_4$ sheet through \eqref{eq:antipode_preserves_P4}. A state specified only by $P_\mu$ may therefore be expanded into branch-resolved components, and the constrained two-particle state takes the auxiliary branch form \eqref{eq:auxiliary_expanded_state}.

This should be read with some care. The branch structure is not a basis-independent statement about arbitrary coordinates on $\kappa$-Poincar\'e momentum space. It is a realization-dependent feature of the classical-basis map from the ordered-plane-wave variables to the variables $P_\mu$. If $P_\mu$ are the directly measured quantities, the main physical content is the deformed correlation pattern implied by the coproduct, including the fact that information invisible in the one-particle projected label can be recovered from branch-sensitive two-particle correlations. If the auxiliary labels $p_\mu$ are themselves operationally meaningful, the same constrained state can be interpreted as a superposition of different auxiliary branches associated with the same $P_\mu$. In either interpretation, the covariance of the construction follows from the Hopf-algebra homomorphism property, as shown in Appendix~\ref{sec:lorentz_covariance}; the correlations are not an artifact of a particular inertial observer.

We also compared this mechanism with standard nonrelativistic minimal-length models, including polymer quantum mechanics and GUP realizations. In regular self-adjoint domains where the deformed commutator remains positive and nonsingular, the corresponding one-dimensional auxiliary map is monotonic and therefore does not admit a smooth local two-real-branch inverse. Periodic completions such as the sine or tangent maps can have global multivaluedness, and singular models can have algebraic extra roots outside the admissible domain, but neither reproduces the local high-momentum two-branch structure found here. The mechanism is therefore genuinely tied to the relativistic energy-momentum mixing and the $P_+$-dependent coproduct of the classical-basis $\kappa$-Poincar\'e framework.

\vspace{0.7cm}

{\bf Acknowledgments:} The work of M.A.G. is supported by IBS under the project code, IBS-R018-D3.
\vspace{0.7cm}

\appendix

\section{Jacobian of the classical-basis map}
\label{sec:jacobian_appendix}

In this appendix we compute in detail the Jacobian of the nonlinear map \eqref{eq:kappa_map} from the ordered-plane-wave labels $p_\mu$ to the classical-basis variables $P_\mu$ and derive the compact expression for its determinant quoted in Eq.~\eqref{eq:jacobian_determinant}.

Starting from
\begin{align}
	P_0 &= \kappa \sinh\!\left(\frac{p_0}{\kappa}\right)
	+ \frac{\vec p^{\,2}}{2\kappa}e^{p_0/\kappa},
	\label{eq:appendix_P0}
	\\
	P_i &= p_i e^{p_0/\kappa},
	\label{eq:appendix_Pi}
\end{align}
we define the Jacobian matrix
\begin{equation}
	J_{\mu\nu}=\frac{\partial P_\mu}{\partial p_\nu}.
\end{equation}
The derivatives of $P_0$ are
\begin{align}
	\frac{\partial P_0}{\partial p_0}
	&=
	\cosh\!\left(\frac{p_0}{\kappa}\right)
	+
	\frac{\vec p^{\,2}}{2\kappa^2}e^{p_0/\kappa},
	\label{eq:appendix_dP0_dp0}
	\\
	\frac{\partial P_0}{\partial p_j}
	&=
	\frac{p_j}{\kappa}e^{p_0/\kappa}.
	\label{eq:appendix_dP0_dpj}
\end{align}
The derivatives of $P_i$ are
\begin{align}
	\frac{\partial P_i}{\partial p_0}
	&=
	\frac{p_i}{\kappa}e^{p_0/\kappa},
	\label{eq:appendix_dPi_dp0}
	\\
	\frac{\partial P_i}{\partial p_j}
	&=
	e^{p_0/\kappa}\delta_{ij}.
	\label{eq:appendix_dPi_dpj}
\end{align}

It is convenient to write the resulting $4\times 4$ Jacobian matrix in block form:
\begin{equation}
	J=
	\begin{pmatrix}
		A & B^T\\
		B & e^{p_0/\kappa}\,\mathbb{I}_3
	\end{pmatrix},
\end{equation}
where
\begin{equation}
	A=
	\cosh\!\left(\frac{p_0}{\kappa}\right)
	+
	\frac{\vec p^{\,2}}{2\kappa^2}e^{p_0/\kappa},
	\qquad
	B_i=\frac{p_i}{\kappa}e^{p_0/\kappa}.
\end{equation}
Since the lower-right block is invertible for all real $p_0$, we may use the standard block-determinant formula
\begin{equation}
	\det J
	=
	\det\!\bigl(e^{p_0/\kappa}\mathbb{I}_3\bigr)\,
	\det\!\left(
	A
	-
	B^T\bigl(e^{p_0/\kappa}\mathbb{I}_3\bigr)^{-1}B
	\right).
\end{equation}
Now
\begin{equation}
	\det\!\bigl(e^{p_0/\kappa}\mathbb{I}_3\bigr)=e^{3p_0/\kappa},
\end{equation}
and
\begin{equation}
	\bigl(e^{p_0/\kappa}\mathbb{I}_3\bigr)^{-1}=e^{-p_0/\kappa}\mathbb{I}_3.
\end{equation}
Therefore
\begin{align}
	B^T\bigl(e^{p_0/\kappa}\mathbb{I}_3\bigr)^{-1}B
	&=
	e^{-p_0/\kappa}\sum_i B_i^2
	\nonumber\\
	&=
	e^{-p_0/\kappa}\sum_i
	\left(
	\frac{p_i}{\kappa}e^{p_0/\kappa}
	\right)^2
	\nonumber\\
	&=
	\frac{\vec p^{\,2}}{\kappa^2}e^{p_0/\kappa}.
\end{align}
Hence
\begin{align}
	A-B^T\bigl(e^{p_0/\kappa}\mathbb{I}_3\bigr)^{-1}B
	&=
	\cosh\!\left(\frac{p_0}{\kappa}\right)
	+
	\frac{\vec p^{\,2}}{2\kappa^2}e^{p_0/\kappa}
	-
	\frac{\vec p^{\,2}}{\kappa^2}e^{p_0/\kappa}
	\nonumber\\
	&=
	\cosh\!\left(\frac{p_0}{\kappa}\right)
	-
	\frac{\vec p^{\,2}}{2\kappa^2}e^{p_0/\kappa}.
\end{align}
Substituting this back, we obtain
\begin{equation}\label{eq:appendix_det_intermediate}
	\det J
	=
	e^{3p_0/\kappa}
	\left[
	\cosh\!\left(\frac{p_0}{\kappa}\right)
	-
	\frac{\vec p^{\,2}}{2\kappa^2}e^{p_0/\kappa}
	\right].
\end{equation}

Using the definition
\begin{equation}
	P_4
	=
	\kappa \cosh\!\left(\frac{p_0}{\kappa}\right)
	-
	\frac{\vec p^{\,2}}{2\kappa}e^{p_0/\kappa},
\end{equation}
we may rewrite the bracket in \eqref{eq:appendix_det_intermediate} as $P_4/\kappa$, so that
\begin{equation}
	\det J
	=
	e^{3p_0/\kappa}\frac{P_4}{\kappa}.
\end{equation}
Finally, since
\begin{equation}
	P_+=\kappa e^{p_0/\kappa},
\end{equation}
we have
\begin{equation}
	e^{3p_0/\kappa}=\left(\frac{P_+}{\kappa}\right)^3,
\end{equation}
and therefore
\begin{equation}\label{eq:appendix_det_final}
	\det\!\left(\frac{\partial P_\mu}{\partial p_\nu}\right)
	=
	e^{3p_0/\kappa}\frac{P_4}{\kappa}
	=
	\frac{P_+^3 P_4}{\kappa^4}.
\end{equation}

This proves Eq.~\eqref{eq:jacobian_determinant}. Since $e^{3p_0/\kappa}>0$ for real $p_0$, the Jacobian vanishes if and only if
\begin{equation}
	P_4=0.
\end{equation}
In terms of the auxiliary variables, this condition is
\begin{equation}
	\kappa \cosh\!\left(\frac{p_0}{\kappa}\right)
	-
	\frac{\vec p^{\,2}}{2\kappa}e^{p_0/\kappa}=0.
\end{equation}
Multiplying by $2\kappa$ and using $\cosh x=\frac12(e^x+e^{-x})$, one finds
\begin{align}
	2\kappa^2 \cosh\!\left(\frac{p_0}{\kappa}\right)
	&=
	\vec p^{\,2}e^{p_0/\kappa},
	\nonumber\\
	\kappa^2\left(e^{p_0/\kappa}+e^{-p_0/\kappa}\right)
	&=
	\vec p^{\,2}e^{p_0/\kappa},
	\nonumber\\
	\kappa^2\left(1+e^{-2p_0/\kappa}\right)
	&=
	\vec p^{\,2}.
\end{align}
Using $\vec P^{\,2}=\vec p^{\,2}e^{2p_0/\kappa}$, this becomes
\begin{equation}
	e^{2p_0/\kappa}
	=
	\frac{\vec P^{\,2}}{\kappa^2}-1.
\end{equation}
This is precisely the same condition that appears in the main text from the turning-point analysis of the inverse problem, confirming that the onset of local multibranch inversion is exactly the point at which the map \eqref{eq:kappa_map} ceases to be locally invertible.

\section{Lorentz covariance of coproduct-induced two-particle states}
\label{sec:lorentz_covariance}

In this appendix, we show, in a fully algebraic way, that the two-particle states defined by the classical-basis coproduct constraint transform covariantly under $\kappa$-deformed Lorentz transformations, using only the Hopf-algebraic structure of $\kappa$-Poincar\'e symmetry.

We work in the classical basis of the $\kappa$-Poincar\'e algebra.
The action of the boost generators $N_i$ on the translation generators
$P_\mu$ is given by \eqref{eq:boost_energy_action} and \eqref{eq:boost_momentum_action}.
These relations define the ordinary Lorentz action on the classical-basis four-vector $P_\mu$.
The finite boost transformation of momenta is generated by
\begin{equation}\label{eq:finite_boost}
	P_\mu \rightarrow P_\mu' = e^{i \xi^i N_i} \, P_\mu \, e^{-i \xi^i N_i},
\end{equation}
where $\xi^i$ denotes the rapidity parameters. A crucial consistency requirement is that the coproduct
$\Delta(P_\mu)$ transforms covariantly under boosts.
The relevant Hopf-algebra data are \eqref{eq:coproduct_energy}--\eqref{eq:rotation_coproduct}, in particular
\begin{equation}\label{eq:boost_coproduct_appendix}
	\Delta(N_i)
	=
	N_i \otimes 1
	+
	\frac{\kappa}{P_+}\otimes N_i
	+
	\sum_{j,k}\frac{1}{P_+}\epsilon_{ijk} P_j \otimes M_k .
\end{equation}
The covariance condition is expressed as
\begin{equation}\label{eq:coproduct_covariance}
	\Delta\!\left( [N_i, P_\mu] \right)
	=
	[\Delta(N_i), \Delta(P_\mu)].
\end{equation}
This equation follows from the fact that
the coproduct $\Delta$ is an algebra homomorphism,
ensuring the consistent action of symmetry generators
on composite systems. Using \eqref{eq:coproduct_energy}--\eqref{eq:rotation_coproduct}, one verifies directly that
\begin{equation}\label{eq:covariance_components}
	[\Delta(N_i),\Delta(P_0)] = i\,\Delta(P_i),
	\qquad
	[\Delta(N_i),\Delta(P_j)] = i\,\delta_{ij}\,\Delta(P_0),
\end{equation}
and similarly for rotations. Equivalently, at the finite level one has
\begin{equation}\label{eq:finite_coproduct_covariance}
	e^{i \xi^i \Delta(N_i)}\,\Delta(P_\mu)\,e^{-i \xi^i \Delta(N_i)}
	=
	\Delta(P'_\mu).
\end{equation}
Therefore, the deformed momentum composition encoded in $\Delta(P_\mu)$
is fully compatible with deformed Lorentz transformations.

We now examine the behavior of the total momentum constraint \eqref{eq:total_momentum_zero} under a Lorentz boost.
Applying a finite boost transformation to the state yields
\begin{equation}\label{eq:boosted_state}
	|\Psi\rangle \rightarrow |\Psi'\rangle = e^{i \xi^i \Delta(N_i)} |\Psi\rangle .
\end{equation}
Using Eq.~\eqref{eq:finite_coproduct_covariance}, we obtain
\begin{align}
	\Delta(P'_\mu) |\Psi'\rangle
	&=
	e^{i \xi^i \Delta(N_i)}\,\Delta(P_\mu)\,e^{-i \xi^i \Delta(N_i)}
	e^{i \xi^i \Delta(N_i)} |\Psi\rangle \nonumber \\
	&=
	e^{i \xi^i \Delta(N_i)} \Delta(P_\mu) |\Psi\rangle \nonumber \\
	&= 0 .
	\label{eq:constraint_invariance}
\end{align}
Hence, if the original state satisfies the total momentum constraint
$\Delta(P_\mu)|\Psi\rangle=0$, then the boosted state satisfies the transformed constraint
$\Delta(P'_\mu)|\Psi'\rangle=0$.
Thus, the constraint surface is covariant under deformed Lorentz boosts.
This result ensures that the physical content of the state
is observer independent.

For a branch-resolved constrained two-particle state of the form
\begin{equation}
	|\Psi_k\rangle
	=
	|P^{(1)},k\rangle_1 \otimes \big|P^{(2)}(P^{(1)},k),\ell(k)\big\rangle_2 ,
\end{equation}
with the correlated partner fixed by \eqref{eq:deformed_antipode},
the boosted state is defined by the two-particle representation
\begin{equation}
	|\Psi_k'\rangle
	=
	e^{i \xi^i \Delta(N_i)}
	\Bigl(|P^{(1)},k\rangle_1 \otimes \big|P^{(2)}(P^{(1)},k),\ell(k)\big\rangle_2\Bigr) .
\end{equation}
In general, because the boost coproduct \eqref{eq:boost_coproduct_appendix} is nonprimitive, this transformed state need not factorize into a tensor product of independently boosted one-particle states. What is guaranteed by the Hopf-algebraic covariance established above is that $|\Psi_k'\rangle$ remains a well-defined two-particle state satisfying the transformed constraint
\begin{equation}
	\Delta(P'_\mu)|\Psi_k'\rangle=0 .
\end{equation}

Now consider the coarse-grained state expanded in the auxiliary basis as in Eq.~\eqref{eq:auxiliary_expanded_state}:
\begin{equation}
	|\Psi\rangle
	=
	\sum_k \alpha_k\,
	|p_k\rangle_1\otimes|\tilde p_k\rangle_2 .
\end{equation}
Under the boost, the transformed state is
\begin{equation}
	|\Psi'\rangle
	=
	e^{i \xi^i \Delta(N_i)}
	\sum_k \alpha_k\,
	|p_k\rangle_1\otimes|\tilde p_k\rangle_2 .
\end{equation}
Since the auxiliary tensor-product basis spans the two-particle Hilbert space, the boosted state can again be expanded in that basis,
\begin{equation}\label{eq:boosted_auxiliary_state}
	|\Psi'\rangle
	=
	\sum_{k,\ell}\gamma_{k\ell}\,
	|p'_k\rangle_1\otimes|\tilde p'_\ell\rangle_2 ,
\end{equation}
for some coefficients $\gamma_{k\ell}$. Thus, although the detailed auxiliary-basis coefficients and the factorization properties of the state need not remain unchanged, the boosted constrained state again admits a well-defined auxiliary-basis expansion. In this sense, the coproduct-induced two-particle correlations and their auxiliary-basis representation transform covariantly under $\kappa$-deformed Lorentz transformations.
	
	\bibliographystyle{JHEPmod}
	\bibliography{refs}
	
\end{document}